# A NEW NON-NEGATIVE MATRIX FACTORIZATION APPROACH FOR BLIND SOURCE SEPARATION OF CARDIOVASCULAR AND RESPIRATORY SOUND BASED ON THE PERIODICITY OF HEART AND LUNG FUNCTION

*Yasaman Torabi[1], Shahram Shirani[1,2], James P. Reilly[1]*

[1] Electrical and Computer Engineering Department, McMaster University, Hamilton, Ontario, Canada
[2] L.R. Wilson/Bell Canada Chair in Data Communications, Hamilton, Ontario, Canada

## ABSTRACT

Auscultation provides a rich diversity of information to diagnose cardiovascular and respiratory diseases. However, sound auscultation is challenging due to noise. In this study, a modified version of the affine non-negative matrix factorization (NMF) approach is proposed to blindly separate lung and heart sounds recorded by a digital stethoscope. This method applies a novel NMF algorithm, which embodies a parallel structure of multilayer units on the input signal, to find a proper estimation of source signals. Another key innovation is the use of the periodic property of the signals which improves accuracy compared to previous works. The method is tested on 100 cases. Each case consists of two synthesized mixtures of real measurements. The effect of different parameters is discussed, and the results are compared to other current methods. Results demonstrate improvements in the source-to-distortion ratio (SDR), source-to-interference ratio (SIR), and source-to-artifacts ratio (SAR) of heart and lung sounds, respectively.

## 1. INTRODUCTION

Cardiovascular disease (CVD) refers to a class of diseases involving the heart and/or blood vessels [1]. In Canada, CVD accounted for 29.8%, or 81,300 deaths per year in 2016 [2]. Moreover, there are abundant cases of respiratory diseases in this country. In Canada, 3.8 million people over the age of one are living with asthma and 2.0 million are living with chronic obstructive pulmonary disease (COPD), both of which can impact a person's ability to breathe [3]. Therefore, accurate and timely assessment for signs of serious health problems such as cardio-respiratory diseases is an essential requirement to provide adequate health care [4].

There are various methods to acquire respiratory and cardiac signals; the phonocardiogram (PCG) represents the recording of sounds, and electrocardiography (ECG) records the heart's electrical activity. ECG is the most popular method for checking cardiac anomalies. Therefore, there have been advances in the design of Holter monitoring devices for obtaining electrocardiography waveforms [5], [6]. However, heart disorders caused by structural abnormalities are more likely to produce mechanical, rather than electrical vibrations [7]. This leads to a better understanding of the importance of cardiac auscultation. Additionally, Internet-of-things (IoT) innovations have led to the development of precise monitoring and remote diagnostics devices [26]-[29].

Auscultation is one of the most fundamental ways to evaluate heart functions. A stethoscope can be used to auscultate respiratory sounds and heart sounds and diagnose most cardiopulmonary disorders and other diseases [8]. However, extraction of desired body sounds without interference from other body sounds is challenging. Various body sounds interact with each other; e.g., cardio, and pulmonary sounds [9]. To achieve accurate recognition, sound separation is an important pre-processing step. Because the measured signal is usually a mixed version of heart and lung sounds, and pure heart/lung acoustic signals are generally not accessible, effectively separating heart and lung sounds is a very challenging prospect. Moreover, the frequency range of the heart and lung sounds can be highly overlapped. This results in interference between the acoustic signals [10].

Blind source separation (BSS) is a technique for separating specific sources from a recorded sound without any information about the recording environment, mixing system, or source locations [11]. Non-negative matrix factorization (NMF) and independent component analysis (ICA) are the mainstream methods for blind source separation. ICA is a generative model to find a linear decomposition of the observed data such that the constituent components are statistically independent, or as independent as possible. On the other hand, NMF is a matrix factorization technique that decomposes a non-negative matrix into a pair of non-negative matrices with a lower rank [12]. Even though the application of NMF for BSS is well studied, there is little work that exploits this method for the separation of bioacoustic signals. In this paper, a new NMF approach for heart and lung sound separation is proposed. The algorithm is assessed using synthesized mixtures of real-world data. Finally, the results are compared with other methods. The experimental results demonstrate a superior source separation accuracy when compared to other methods.

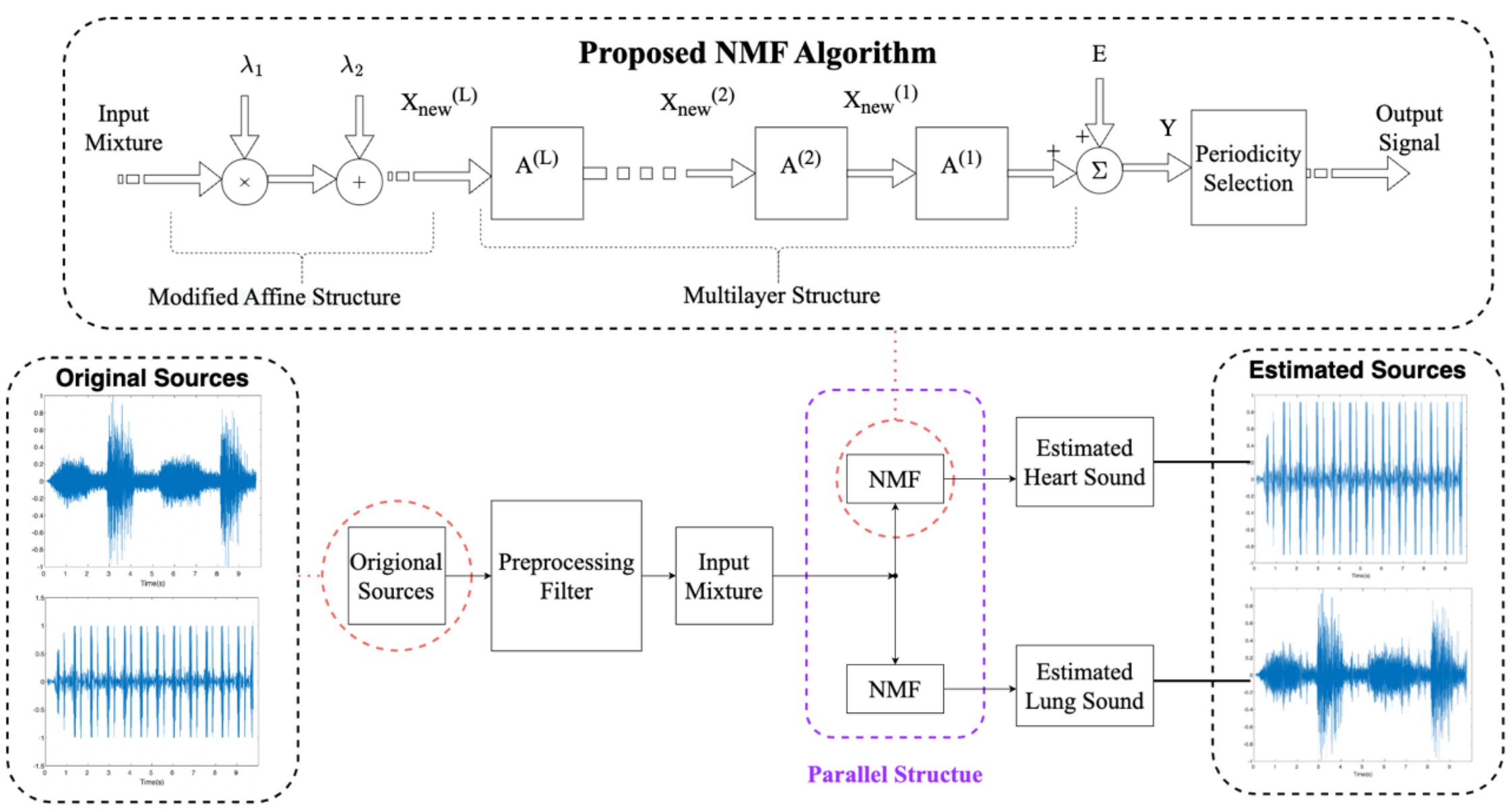

**Fig. 1**: Overview of the stages of the proposed source separation algorithm.

**Related works:** There is little work on the application of NMF for heart and lung sound separation. In the latest work, Grooby et al. [4] adapted the standard NMF method to improve performance. A limitation of the method proposed in [4] is the quality of the lung sound results, which still contain noise and some remains of the heart sound. They manually separated the lung segments, which is not required in our proposed method.

Most current heart and lung sound separations employ methods other than NMF, which normally requires prior knowledge. For instance, ICA is mainly used for this application [13]-[15]. One drawback of using ICA over NMF is that it requires prior knowledge about the mean and variance of the sources. There is also another proposed method for this application. Tsalaile et al. [16] proposed an approach based on a sequential approximate diagonalization algorithm (SDA) exploiting periodicity. The performance of this algorithm depends on prior knowledge of the period of the source signals. In our proposed method, however, no prior knowledge is needed, except the fact that original sources are periodic, no matter what their periods are. The periodicity feature of heart and lung sounds results in improved accuracy. This is a natural feature that has not been used in NMF-based algorithms before, to the best of our knowledge.

In addition, Sheikh et al. [9] exploited multiple signal classification (MUSIC) algorithm for the radial filtering of body Sounds. MUSIC assumes coexistent sources to be uncorrelated that limits its practical applications, but this assumption is not needed in NMF-based methods.

Our proposed algorithm not only is new in terms of application but also has a novel contribution, as explained in the following. This allows the proposed method to be used for other source separation applications as well. The main disadvantage of NMF is its limitation in recovering mixed-sign signals. Our method consists of a scale-and-offset block which is a modification of the affine NMF algorithm [17]. It is capable of handling mixed-sign mixtures of heart and lung sound thanks to its adjustable offset term. Moreover, its scaling feature helps the algorithm to find the original sources with better accuracy, especially if any prior knowledge regarding the amplitude of each original source is given.

A multilayer structure has been previously used for NMF to reduce the risk of converging to local minima [18]. One novel aspect of our proposed algorithm includes its multilayer-parallel structure which considerably improves the performance. The parallel structure provides more degrees of freedom to adjust the parameters of heart sound and lung sound estimation separately, which leads to better results.

## 2. METHODOLOGY

The basic NMF problem can be stated as follows [19]: Given a nonnegative data matrix $Y \in \mathbb{R}_+^{I\times T}$ (with $Y \geq 0$), find two nonnegative matrices $A = [a_1, a_2, \ldots, a_J] \in \mathbb{R}_+^{I\times J}$ and $X = [x_1, x_2, \ldots, x_J] \in \mathbb{R}_+^{J\times T}$ which factorize Y such that:

$$Y = AX + E \tag{1}$$

where the matrix $E \in \mathbb{R}^{I\times T}$ represents approximation error, A is the mixing signal and X is the source signal.

In this paper, we use $\alpha$-NMF algorithm. X and A are randomly initiated. Then, their values are updated by minimizing the cost function, which is α-divergence distance described as follows:

$$D_A^{\alpha}(Y||AX) = \frac{1}{\alpha(\alpha-1)}\sum_{it}\left(y_{it}^{\alpha}[AX]_{it}^{1-\alpha} - \alpha y_{it} + (\alpha-1)[AX]_{it}\right) \quad (2)$$

The update rules are as follows:

$$x_{jt} \leftarrow x_{jt}\left(\sum_{i\in S_I}\widehat{a_{ij}}\left(\frac{y_{it}}{[AX]_{it}}\right)^{\alpha}\right)^{\frac{1}{\alpha}} \quad (3)$$

$$a_{ij} \leftarrow a_{ij}\left(\sum_{t\in S_T}x_{jt}\left(\frac{y_{it}}{[AX]_{it}}\right)^{\alpha}\right)^{\frac{1}{\alpha}} \quad (4)$$

The algorithm is repeated until the stopping criterion is met, which is as follows:

$$\frac{\left|D_f^{(k)} - D_f^{(k-1)}\right|}{D_f} \leq \epsilon \quad (5)$$

Fig. 1 shows an overview of the different stages of the method, how they are connected, and what they are used for.

### 2.1. Pre-Processing

We surpass the noise by applying two bandpass filters: 50-250 Hz for the heart signal [20], and 60-300 Hz for the lung signal [21]. The filters have minimum order with a stopband attenuation of 60 dB. We recorded original sources and thus filtered them before mixing. In cases where mixtures are recorded, the filter is applied to mixtures. Moreover, we normalize signals as follows:

$$x_{normalized} = \frac{x - \mu}{max(|x|)} \quad (6)$$

Where $\mu$ is the mean of x.

### 2.2. Modified Affine NMF

In standard affine NMF, the goal is to remove the baseline, increase sparsity, and improve uniqueness [19]. With ordinary affine NMF, an offset is applied to the input signal. As a novel contribution, we propose a modified transformation as shown in Fig. 1 by adding an extra scaling factor to increase the performance. Some biological prior knowledge can lead to better parameter selection. Lung sounds change more slowly, and they are more dominant than heart sounds. Therefore, we choose two separate scaling factors, $\lambda_{1\,lung} \epsilon (0,1)$ for the lung-detection block and $\lambda_{1\,heart} \geq 1$ for the heart-detection block. Furthermore, we employ an offset value $\lambda_2$ that is chosen such that the final signal is nonnegative and as close as possible to zero baseline as follows:

$$\lambda_1 \times min(signal) + \lambda_2 \geq 0 \quad (7)$$

### 2.3. Multilayer NMF

The hierarchical multi-stage NMF procedure considerably improves the performance and reduces the risk of converging to local minima. In multi-layer NMF, the basic matrix A is replaced by a set of cascaded matrices [19]. Thus, the model can be described as:

$$Y = A^{(1)}A^{(2)} \ldots A^{(L)}X + E \quad (8)$$

### 2.4. Periodicity-based Parallel Analysis

All separation algorithms exploit one NMF module to separate the mixtures, to our best knowledge. Our proposed method applies two separate NMF modules on the input, as shown in Fig. 1. Each module generates two signals, and the unwanted signal from each module is disregarded such that one module extracts the heart sound and the other extracts the lung sound. This is accomplished using the periodic nature of heart and lung sounds. There is no need to have prior knowledge of the approximate period of each source. We estimate the period of each output at the end of the process and compare them. We evaluate the periodicity of the signal using the period estimation of the autocorrelation function. One module chooses the output with a smaller period, which is the estimation of heart sound, and the other module selects the lung sound. This novel approach provides more degrees of freedom for parameter selection and lets us tune the parameters of each module (i.e., $\lambda_1$, $\lambda_2$, $\alpha$, $L$) separately, which leads to better accuracy. Estimation of each source with a fixed NMF module improves the average performance of samples.

### 2.5. Evaluation Criteria

We normalize all signals so that they are comparable. We use the available pure heart and lung sounds as a reference to compute the separation performance and use source-to-distortion ratio (SDR), source-to-interference ratio (SIR), and source-to-artifacts ratio (SAR). Larger values refer to better separation [22]. The estimated source can be decomposed as follows:

$$\hat{s} = s^{target} + e^{interference} + e^{noise} + e^{artifact} \quad (9)$$

Where $s$ refers to source signal and $e$ refers to unwanted signals. Then, the evaluation criteria are defined as:

$$SDR = 10log_{10}\left(\frac{||s^{target}||_2^2}{||e^{interference} + e^{noise} + e^{artifact}||_2^2}\right) \quad (10)$$

$$SIR = 10log_{10}\left(\frac{||s^{target}||_2^2}{||e^{interference}||_2^2}\right) \quad (11)$$

$$SAR = 10log_{10}\left(\frac{||s^{target} + e^{interference} + e^{noise}||_2^2}{||e^{artifact}||_2^2}\right) \quad (12)$$

## 3. EXPERIMENTS AND RESULTS

### 3.1. Dataset

10-second heart and lung sound measurements are recorded using a digital stethoscope in a silent environment. The sources are mixed to create the synthesized dataset of 100 mixtures using MATLAB [30].

### 3.2. Implementation and Setup

We set $\lambda_1$=0.2, $\lambda_2$=1 for heart detection, and $\lambda_1$=5, $\lambda_2$=6 for lung detection. The weights are optimized using gradient descent with 1000 iterations and learning rate=1e-5. The proposed work is implemented in MATLAB.

In order to select optimal parameters, we conduct 100 experiments. Various $\alpha$ and number of layers are tested to find the optimal model based on the mean of all estimated SIRs for each set of parameters. The BSS_EVAL toolbox [23] is applied to measure performance. Based on Table 1, we choose α=0.5 for both NMF blocks. Heart detection block has 2 layers and lung detection block has 1 layer.

**Table 1**: Averaged SIR [dB] of heart and lung sound separation using different parameters after 100 experiments. 'H' stands for heart and 'L' for lung, respectively. Optimal parameters for each source detection are highlighted.

| $\alpha$ | Layer 1 | | Layer 2 | | Layer 3 | | Layer 4 | |
|---|---|---|---|---|---|---|---|---|
| | H | L | H | L | H | L | H | L |
| -1 | 22.4 | 21.8 | 23.7 | 20.7 | 23.2 | 20.4 | 23.1 | 20.0 |
| 0.5 | 22.3 | **32.5** | **29.2** | 31.8 | 28.7 | 27.6 | 28.7 | 27.6 |
| 1 | 22.9 | 31.6 | 28.9 | 29.3 | 28.8 | 25.9 | 28.8 | 25.7 |
| 2 | 24.9 | 30.0 | 28.2 | 26.5 | 27.9 | 23.9 | 27.8 | 23.4 |
| 10 | 21.3 | 24.9 | 21.0 | 20.3 | 20.9 | 18.6 | 20.8 | 18.2 |

### 3.3. Evaluation and Result

Fig. 2 shows the original sources, two example mixtures, and the estimated outputs. Comparative results between the proposed method and recent NMF-based methods are given in Table 2. Compared to several related works. the results show that our method provided satisfactory separation results and achieved superior quality, especially in SIR, which is the most relevant metric for clinical purposes. The results can be enhanced by applying more advanced pre-processing filters in future work. Moreover, the proposed method will be tested on clinical data in further studies, as promising results are obtained on simulations.

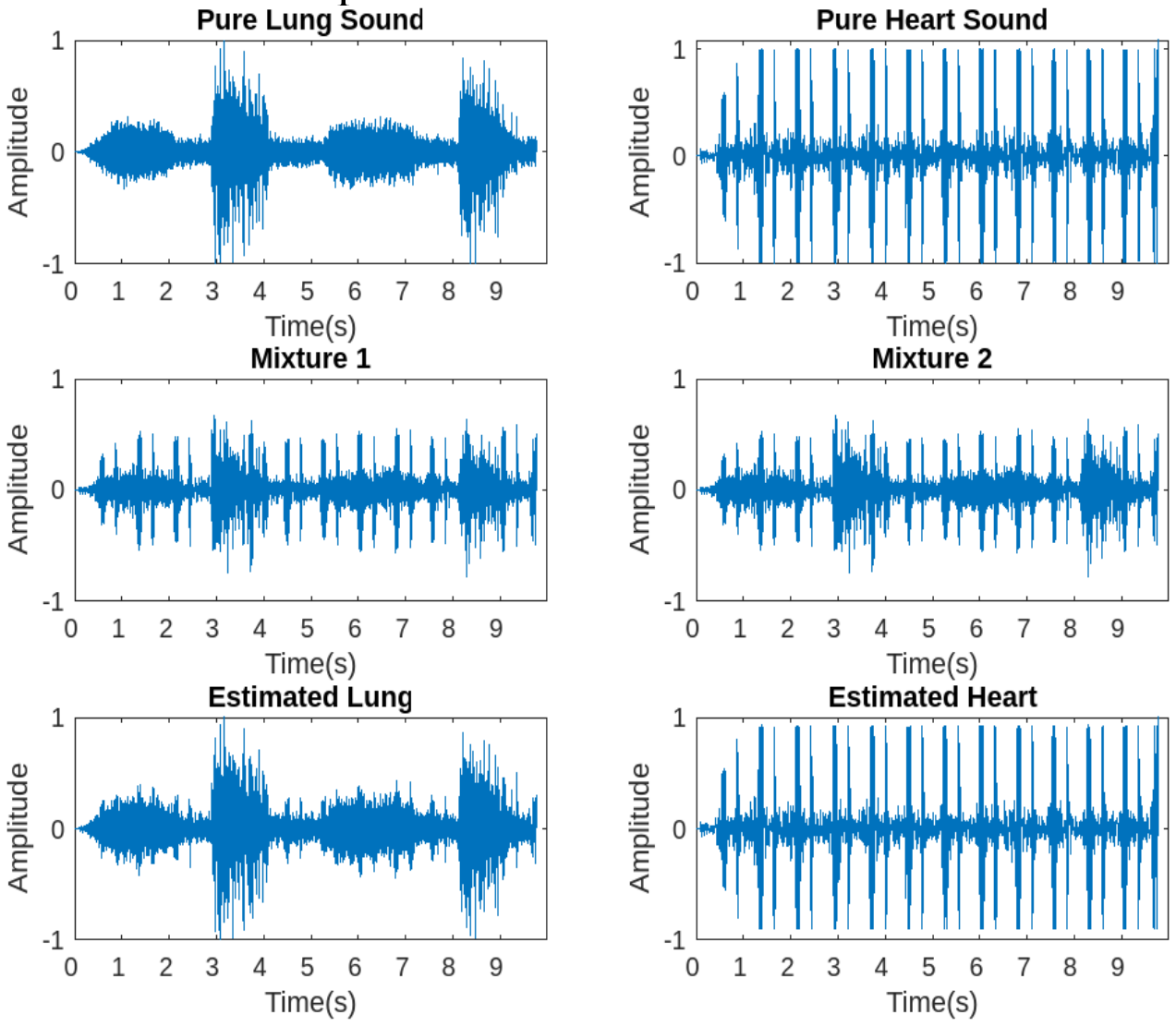

**Fig. 2**: Original sources, two example mixtures, and the estimated outputs.

**Table 2**: Performance comparison of the proposed algorithm and recent NMF-based separation methods.

| Method | Data Type | SIR [dB] | SAR [dB] | SDR [dB] |
|---|---|---|---|---|
| Tsai [10] | Synthetic | 10.4 | 14.9 | 8.7 |
| Xie [22] | Clinical | 14.7 | 8.1 | 7.1 |
| Wang [24] | Clinical | 23.2 | 10.6 | 8.7 |
| Candas [25] | Clinical | 24.1 | 17.9 | 17.3 |
| **Proposed Method** | Synthetic | 30.9 | 17.1 | 21.2 |

## 4. CONCLUSION

In this paper, we proposed a novel algorithm based on non-negative matrix factorization (NMF) for blind source separation of heart and lung sound, combining parallel blocks and multi-layer structures. Besides, the proposed method exploits a prior knowledge of the body signals and thus is based on periodicity to improve existing blind source separation methods. The method was statistically evaluated after executing 100 times with two different mixtures of heart and lung sounds. The experimental results demonstrated that the proposed method was superior to previous methods in heart and lung sound separation.

## 5. REFERENCES

[1] J. Philip, et al., “Reducing the Global Burden of Cardiovascular Disease, Part 1: The Epidemiology and Risk Factors,” *Circulation research,* vol. 121, no. 6, pp. 677-694, 2017.

[2] JJ. Lang, S. Alam, LE. Cahill, et al., “Global burden of disease study trends for Canada from 1990 to 2016,” *CMAJ,* vol 190, no. 44, pp. E1296–304, 2018.

[3] P. H. A. of Canada, “Government of Canada,” *Canada.ca*, Available: https://www.canada.ca/en/public-health/services/publications/diseases-conditions/asthma-chronic-obstructive-pulmonary-disease-canada-2018.html [Accessed: 09-Oct-2022].

[4] E. Grooby, et al., "A New Non-Negative Matrix Co-Factorisation Approach for Noisy Neonatal Chest Sound Separation," *43rd Annual International Conference of the IEEE Engineering in Medicine & Biology Society (EMBC)*, pp. 5668-5673, 2021.

[5] K. Swetha, and T. Jayasree, "Analytical Study of Functional Trends in Cardiorespiratory Activity," *International Conference on Applied Artificial Intelligence and Computing (ICAAIC)*, pp. 1699-1703, 2022.

[6] B. Baraeinejad, M. F. Shayan, A. R. Vazifeh, et al., "Design and Implementation of an Ultralow-Power ECG Patch and Smart Cloud-Based Platform," *IEEE Transactions on Instrumentation and Measurement*, vol. 71, pp. 1-11, no. 2506811, 2022.

[7] H. Ren, et al., “A Novel Cardiac Auscultation Monitoring System Based on Wireless Sensing for Healthcare,” *IEEE journal of translational engineering in health and medicine,* vol. 6, no. 1900312, 2018.

[8] J. Y. Shin, S. L'Yi, D. H. Jo, J. H. Bae and T. S. Lee, "Development of smartphone-based stethoscope system," *13th International Conference on Control, Automation and Systems (ICCAS 2013)*, pp. 1288-1291, 2013.

[9] M. A. Sheikh, L. Kumar and M. T. Beg, "Circular Microphone Array Based Stethoscope for Radial Filtering of Body Sounds," *International Conference on Power Electronics, Control and Automation (ICPECA)*, pp. 1-6, 2019.

[10] K. -H. Tsai, et al., "Blind Monaural Source Separation on Heart and Lung Sounds Based on Periodic-Coded Deep Autoencoder," *IEEE Journal of Biomedical and Health Informatics*, vol. 24, no. 11, pp. 3203-3214, 2020.

[11] K. Daichi, et al., “Determined Blind Source Separation Unifying Independent Vector Analysis and Nonnegative Matrix Factorization,” *IEEE/ACM Transactions on Audio, Speech, and Language Processing*, vol. 24, no. 9, 2016.

[12] A. Mirzal, “NMF versus ICA for blind source separation,” *Advances in Data Analysis and Classification*, pp. 25–48, 2017.

[13] J. -C. Chien, et al., "A Study of Heart Sound and Lung Sound Separation by Independent Component Analysis Technique," *International Conference of the IEEE Engineering in Medicine and Biology Society*, pp. 5708-5711, 2006.

[14] F. Ayari, M. Ksouri, and A. T. Alouani, "Lung sound extraction from mixed lung and heart sounds FASTICA algorithm," *16th IEEE Mediterranean Electrotechnical Conference*, pp. 339-342, 2012.

[15] M. T. Pourazad, Z. Moussavi, F. Farahmand, and R. K. Ward, "Heart Sounds Separation from Lung Sounds Using Independent Component Analysis," *IEEE Engineering in Medicine and Biology 27th Annual Conference*, pp. 2736-2739, 2005.

[16] T. Tsalaile, S. M. Naqvi, K. Nazarpour, S. Sanei and J. A. Chambers, "Blind source extraction of heart sound signals from lung sound recordings exploiting periodicity of the heart sound," *IEEE International Conference on Acoustics, Speech and Signal Processing*, pp. 461-464, 2008.

[17] H. Laurberg, and L. K. Hansen, "On Affine Non-Negative Matrix Factorization," *IEEE International Conference on Acoustics, Speech, and Signal Processing (ICASSP),* pp. II-653-II-656, 2007.

[18] Y. E. Salehani, and S. Gazor, "Sparse data reconstruction via adaptive ℓp-norm and multilayer NMF," *IEEE 7th Annual Information Technology, Electronics and Mobile Communication Conference (IEMCON)*, pp. 1-6, 2016.

[19] A. Cichocki, et al., “Nonnegative matrix and tensor factorizations-Applications to exploratory multi-way data analysis and blind source separation,” *Wiley*, 2009.

[20] S. R. Thiyagaraja, et al., ‘‘A novel heart-mobile interface for detection and classification of heart sounds,’’ *Biomed. Signal Process. Control*, vol. 45, pp. 313–324, 2018.

[21] M. Bahoura, ‘‘Analyse des signaux acoustiques respiratoires: Contribution à la dØtection automatique des sibilants par paquets d’ondelett,” *Doctoral thesis in medical sciences,* Rouen, 1999.

[22] Xie, et al., “Reverberant blind separation of heart and lung sounds using nonnegative matrix factorization and auxiliary function technique,” *Biomedical Signal Processing and Control*, vol. 69, no. 102899, 2021.

[23] C. F´evotte, R. Gribonval, and E. Vincent, “BSS EVAL Toolbox User Guide,” *IRISA Technical Report 1706*, Rennes, France, 2005.

[24] W. Wang, et al., “Heart-Lung Sound Separation by Nonnegative Matrix Factorization and Deep Learning,” *Biomedical Signal Processing and Control*, Vol. 79, no. 104180, 2023.

[25] F.J. Canadas-Quesada, et el., “A non-negative matrix factorization approach based on Spectro-temporal clustering to extract heart sounds,” *Applied Acoustics*, Vol. 125, pp. 7-19, 2017.

[26] Y. Torabi et al., "Exploring Sensing Devices for Heart and Lung Sound Monitoring," arXiv preprint, doi: 2406.12432, 2024.

[27] Torabi, Y., Shirani, S., Reilly, J. P., & Gauvreau, G. M. (2024). MEMS and ECM Sensor Technologies for Cardiorespiratory Sound Monitoring-A Comprehensive Review. Sensors (Basel, Switzerland), 24(21), 7036.

[28] Baraeinejad, B., Shams, M., Hamedani, M. S., et al. (2023). Clinical IoT in Practice: A Novel Design and Implementation of a Multi-functional Digital Stethoscope for Remote Health Monitoring. TechRxiv. doi: 10.36227/techrxiv.24459988.v1

[29] Baraeinejad, B., Forouzesh, M., et al. (2024). Design and Implementation of an IoT-based Respiratory Motion Sensor. arXiv preprint arXiv:2412.05405.

[30] Torabi, Y., Shirani, S., & Reilly, J. P. (2024). Manikin-Recorded Cardiopulmonary Sounds Dataset Using Digital Stethoscope. arXiv preprint arXiv:2410.03280.